\documentclass[onecolumn,useAMS]{mn2e}
\usepackage{epsfig}
\def\la{\lower.5ex\hbox{$\; \buildrel < \over \sim \;$}}
\def\ga{\lower.5ex\hbox{$\; \buildrel > \over \sim \;$}}
\def\apj{ApJ}
\def\mnras{MNRAS}
\def\prd{PRD}

\def \be{\begin{equation}}
\def \ee{\end{equation}}
\def \bea{\begin{eqnarray}}
\def \eea{\end{eqnarray}}
\def\etal{{et al.\ }}
\begin{document}
\title[Primordial magnetic fields and formation of molecular hydrogen]{Primordial magnetic fields and formation of molecular hydrogen}
\author[Shiv K. Sethi, Biman B. Nath, Kandaswamy Subramanian]
{Shiv K. Sethi$^1$, Biman B. Nath$^1$, Kandaswamy Subramanian$^2$ \\
\hspace{-0.1cm} $^1$Raman Research Institute, Sadashivanagar, Bangalore 560080, India\\
\hspace{-0.1cm} $^2$Inter-University Center for Astronomy \& Astrophysics, Post Bag 4, Ganeshkhind, Pune 411007, India\\
\hspace{-0.1cm} emails: sethi, biman@rri.res.in. kandu@iucaa.ernet.in
}
\maketitle
\begin{abstract}
We  study the implications of primordial magnetic fields 
for the thermal and ionization history of the post-recombination
era. In particular we compute the effects of dissipation of 
primordial magnetic fields owing to ambipolar diffusion and 
decaying turbulence in the intergalactic medium (IGM) and the collapsing halos and compute 
the effects of the altered thermal and ionization history on the formation 
of molecular hydrogen. We show that, for magnetic field 
strengths in the 
range $2 \times 10^{-10} \, {\rm G} \la B_0 \la 2 \times 10^{-9} \, {\rm G}$, 
the molecular hydrogen fraction in IGM and collapsing halo can increase
by a factor 5 to 1000 over the case with no magnetic fields.
 We discuss the implication of the
increased molecular hydrogen fraction on the radiative transfer of UV 
photons and the formation of first structures in the universe. 
\end{abstract}


\section{Introduction}
It is believed that the first luminous objects in the universe
appeared at $z\sim 15\hbox{--}30$ when substantial amount of molecular
hydrogen formed in objects of mass $\simeq  10^{6\hbox{--}7}$ M$_{\odot}$
(Tegmark \etal 1997). 
The first calculations of the production of H$_2$ by Saslaw \& Zipoy (1967)
had proposed that 
 H$_2$ molecules formed through H$_2^+$ as an intermediate
state, but Peebles \& Dicke (1968) and Hirasawa \etal (1969) suggested
that a more dominant intermediate state was the H$^-$ ion. 
Subsequent calculations
by Lepp \& Shull (1984) showed that the intergalactic medium (IGM) would have 
a final H$_2$ abundance of $\simeq  10^{-6}$. Recently, Hirata \&
Padmanabhan (2006) showed the dominance of the H$^-$ intermediate stage
and calculated a final abundance of H$_2 \simeq 6 \times 10^{-7}$ 
in the intergalactic gas for the standard cosmology.

These calculations depend crucially on the residual ionization fraction
of the universe after recombination. Recently, several authors have
considered the effects of a primordial magnetic field, of order
$\simeq  10^{-9}$ G (in comoving frame), on the formation of large scale
structure in the universe (Sethi \& Subramanian 2005; Yamazaki \etal 2006a). 
The dissipation of  primordial magnetic
field energy  is likely to raise the IGM temperature to $\simeq 10^4$ at redshifts
$z \ge 100$, and collisional ionization in this gas is likely to increase
the ionization fraction of the IGM much above the standard level of 
$\simeq  10^{-4}$ assumed in the calculations (of H$_2$ production) mentioned
above (Sethi \& Subramanian 2005).
Here we consider the detailed chemistry of the production of H$_2$ molecule
in the presence of dissipation of primordial magnetic field in the IGM.
We also study the effect of magnetic field
dissipation during the formation of H$_2$
molecules in the collapse of the first objects.

It is
believed that these first luminous
objects were instrumental in ushering the epoch of reionization, through
UV photons radiating from the first generation stars in these objects. 
We know that the universe was reionized by $z\sim 6$ from Gunn-Peterson
test of QSO absorption lines (see, e.g.,
Fan, Carilli \& Keating 2006). Study of the anisotropy of cosmic microwave
background (CMB) has put limits on the electron scattering optical depth 
in the IGM, which shows that the universe was reionized at
$z \simeq  10$ (Spergel \etal 2006).

The presence of magnetic fields in the early universe can significantly
change the standard history of star formation in the universe (Sethi and Subramanian 2005). They also showed that magnetic fields
can seed early formation of structures, with collapse redshift 
almost independent of the strength of the magnetic field. Tashiro
\& Sugiyama (2005) considered several models of 
early reionization  with primordial magnetic fields
 with magnetic fields strengths  $\sim 10^{-9} \, \rm G$. 

In this paper, we consider other important  implications of 
primordial magnetic fields  for the formation of first structures:  (a)
thermal and ionization history for collapsing haloes, and (b) the 
formation of molecular hydrogen. 

In the next section, we discuss the relevant equations. In \S 3, we discuss
the dissipation mechanisms of the magnetic field energy. In \S 4, we present our 
results. The implications of the results for the formation of the first 
structures is discussed in \S 5. We summarize and conclude our 
findings in \S 6. Throughout this paper we 
use  the  FRW  model favoured by the WMAP results (Spergel et al. 2007): 
spatially flat
with $\Omega_m = 0.3$ and $\Omega_\Lambda = 0.7$ 
 with  $\Omega_b h^2 = 0.022$  and 
$h = 0.7$ (Freedman  et al. 2001).

\section{Formation of molecular hydrogen}
In the absence of dust grains, the formation of molecular hydrogen progresses
through two channels--- with either H$^-$ or H$_2^+$ in the intermediate stage.
In the first process the catalyst is an electron, which attaches to a neutral
hydrogen atom radiatively to form H$^-$,
\be
H+e^- \leftrightarrow H^-+\gamma,
\label{eq:recomb}
\ee
which goes to form $\rm H_2$ molecule  by  the  reaction:
\be
H^-+H \rightarrow H_2+e^- \,.
\label{eq:hminus}
\ee
The  $\rm H^-$ ion  can be destroyed (or used up to produce something other than
H$_2$) in two ways. It can be destroyed by energetic photons, mostly 
provided by the cosmic microwave background in the cosmological setting:
\be
H^-+ \gamma \rightarrow H+e^-\,.
\label{eq:cmb}
\ee
It can also combine with a proton in the following way:
\be
H^-+H^+ \rightarrow 2H \,.
\label{eq:mutual}
\ee
The alternative channel through the formation of H$_2^+$ occurs when 
a proton acts as a catalyst:
\be
H+H^+ \leftrightarrow H_2^+ +\gamma \,, \qquad\qquad
 H_2^+ + H \rightarrow H_2+H^+ \,.
\ee
There is also a third channel through the formation of $HeH^+$. Recently,
Hirata \& Padmanabhan have shown that the H$^-$ channel dominates the 
production of the $\rm H_2$ molecule, with only
$\sim 1$\% contribution from the $\rm H_2^+$ channel, and $\sim 0.004$\% from
the $HeH^+$ channel. We therefore use the H$^-$ channel in our calculations,
and the rate of formation of H$^-$ is determined by,
\bea
{d x[H^-] \over dt}&=&k_1 x_e x_{\rm \scriptscriptstyle HI} n_b - k_{\gamma} x[H^-]-k_2 x[H^-] x_{\rm \scriptscriptstyle HI} n_b
\nonumber\\
&& -k_3 x[H^-] x[H^+] n_b \,,
\eea
Here $x_e$, $x_{\rm \scriptscriptstyle HI}$, $x[H^-]$  are the ionized 
fraction, the neutral 
fraction, and the fraction of H-ion $H^-$, respectively; 
 $n_b$ is the total baryon density; $k_1$ is the  rate for the reaction  \ref{eq:recomb}, 
$k_2$ is the rate of formation of $\rm H_2$ in reaction \ref{eq:hminus},
$k_3$ and $k_{\gamma}$ are the rates of destruction of H$^-$ in reactions
\ref{eq:mutual} and \ref{eq:cmb}. We use the fits in Stancil \etal (1998)
for the first three rates:
\bea
k_1 &=& 3 \times 10^{-16} \, \left({T \over 300 \, {\rm K}} \right)^{0.95} \exp(-T/9320 \, {\rm K}) \, {\rm cm}^3 \, 
{\rm s}^{-1} \,, \nonumber\\
k_2 &=& 1.5 \times 10^{-9} \, \left({T \over 300 \, {\rm K}} \right)^{-0.1} \, {\rm cm}^3 \,
{\rm s}^{-1} \,, \nonumber\\
k_3 &=& 4 \times 10^{-8} \, \left({T \over 300 \, {\rm K}} \right)^{-0.5} \, {\rm cm}^3 \,
{\rm s}^{-1} \,.
\eea   
Following Hirata \& Padmanabhan (2007), the rate of destruction by photons
is taken as (for a blackbody spectrum of temperature $T_{\rm cbr}$),
\be
k_{\gamma} (T_{\rm cbr})=  4 \Bigl ( {m_e k_B T_{\rm cbr} 
\over 2 \pi \hbar ^2} \Bigr
) ^{3/2} \, \exp\left(-0.754 \, {\rm eV}/(k_B T_{\rm cbr})\right) \,
 \times k_1 (T_{\rm cbr}) \,,
\ee
where $0.754$ eV is the photo-dissociation threshold energy for the  $\rm H^-$ ion.
We neglect the changes in the rate produced by any distortion of the
microwave background radiation from a pure blackbody spectrum.
These rates are faster than the Hubble rate of expansion for relevant
epochs, so that one can write the steady-state abundance of H$^-$ as,
\be
x[H^-]={k_1 x_e x_{HI} n_b \over k_2 x_{HI} n_b + k_\gamma + k_3 x[H^+] n_b } \,.
\ee

Molecular hydrogen can be destroyed by electrons if the gas is significantly
hotter than $\simeq  10^4$ K;
\be
H_2+e^- \rightarrow H+H^- \,,
\ee 
We include this reaction since our calculation involves gas at high temperatures, 
with a rate adopted from 
Hirasawa (1969),
\be
k_4=2.7 \times 10^{-8} \, T^{-3/2} \, \exp(-43,000 \, {\rm K}/T) \, {\rm cm}^3 \,
{\rm s}^{-1} \,.
\ee
The rate of formation of molecular hydrogen is then given by,
\be
{d x[H_2] \over dt}=k_m n_b x_e (1-x_e-2 x[H_2]) - k_4 n_b x_e x[H_2] \,.
\label{formrate}
\ee
where the rate of formation through the H$^-$ stage is given by,
\be
k_m= {k_1 k_2  x _{HI} n_b \over k_2 x_{HI} n_b + k_\gamma +
k_3 x[H^+] n_b } \,.
\ee
The first term in Eq~(\ref{formrate}) refers to the
net rate of  formation of $H_2$
 through the $H^-$ channel. The second 
 term of  Eq~(\ref{formrate}) refers to the direct 
destruction of $H_2$ by electrons
 with rate coefficient $k_4$.

These equations must be supplemented by equations governing the density
$n_b$
and temperature $T$ of the gas. For the case of IGM expanding with the
Hubble rate, the particle density $n_b$  is given by $n_b=n_0 (1+z)^3$,
where $n_0=\rho_c (0) \Omega_b/(\mu m_p)$, where $\rho_c (0)$ is the
critical density at present epoch, 
$\mu \simeq  1.22$ is the mean molecular weight for a fully neutral  gas (the value 
relevant for us in this paper)
and $m_p$ is proton mass.

The equation determining the gas temperature is given by,
\be
{dT \over dt}={2 \over 3} {\dot{n_b} \over n_b} T+k_{iC} x_e (T_{\rm cbr} -T)
+{2 \over 3 n_b k_B} (L_{heat} -L_{cool}) \,,
\label{temp}
\ee
where the first term describes adiabatic cooling (or heating), the second
term refers to the inverse Compton cooling (or Compton heating), and
$L_{heat}$, $L_{\rm cool}$ are other heating and cooling (volume) rates. We use,
\begin{equation}
k_{iC} = {4 \pi^2 \over 15}  \Bigl ( {kT_{\rm cbr}
\over \hbar c} \Bigr )^4  \, {\hbar \sigma_T 
\over m_e}= 2.48 \times 10^{-22} \, 
T_{\rm cbr}^4 \,\,\,
{\rm s}^{-1}.
\end{equation}
The other important source of cooling apart from the Compton cooling 
is the HI line cooling with (e.g. Cen 1992):
\begin{equation}
L_{\rm cool}  =   7.9\times10^{-19}\left(1+\left({T\over 10^5 \, {\rm K}} \right )^{0.5}\right)^{-1}\times \exp(-118348/T)x_ex_{\rm \scriptscriptstyle HI}n_b \, \, {\rm erg \, s^{-1}}
\end{equation}
The HI line cooling is important when the temperature of the collapsing 
halo exceeds temperatures $\ga 10^4 \, \rm K$. The heating term
in Eq.~(\ref{temp}) 
is given by the dissipation of primordial magnetic field energy  and is 
discussed in the next section. 

The evolution of the ionized fraction is given by:
\begin{equation}
\dot x_e  =    \left [ \beta_e (1-x_e)\exp\left(-h\nu_\alpha/(k_BT_{\rm cbr})\right )-\alpha_e n_b x_e^2 \right ] C + \gamma_e n_b (1-x_e)x_e
\label{eq:elect}
\end{equation}
In this equation, the first two terms are the  terms for the 
recombination of the primeval plasma (for details and notation
 see Peebles 1968, Peebles 1993). For 
$z \la 800$,  $C \simeq 1$ and the first term on the right hand of the equation
rapidly decreases. After the recombination is completed,  the 
only important term is the recombination term (the second term on the right
hand side) which gives a slow decrease in the ionization fraction. The third
term on the right hand side of the equation corresponds to the collisional ionization 
of the IGM. This term is expected to play an important role in 
partially ionizing the IGM owing to the dissipation of magnetic field energy since 
the temperature may become comparable to $10^4 \, \rm K$ in may cases.

For an overdense region collapsing to virialise at a certain redshift 
$z_{vir}$, we calculate the density from the equation of motion of
a bound shell collapsing under gravity (for details see e.g. Padmanabhan 1993, Peebles 1980): 
\be
\ddot{r}=-{GM \over r^2}
\ee
(ignoring the effect of cosmological constant at high redshift). The
parametric solutions of this equation are given by:
\bea
r&=&2r_{\rm vir} [(1-\cos
\theta )/2]\,, \nonumber\\
 t&=&t_c[(\theta-\sin \theta)/2 \pi] \,,.
\eea
Here, $t_c$ is the age of the universe at the
collapse redshift $z_{vir}$, and $r_{\rm vir}=r_{\rm max}/2
={1 \over 2} [(2 G M t_c^2)/\pi ^2]^{1/3}$
is the radius at virialization. The density is assumed to be constant
after virialization, given by the mean overdensity $\sim 18 \pi^2$
 expected in the
spherical top-hat model. In this case, the gas density first
decreases due to universal expansion, and then slowly increases prior to the
collapse redshift to reach an overdensity of $18 \pi^2$ at $z_{vir}$.
The corresponding increase in temperature from adiabatic compression is 
given by the first term of equation \ref{temp}. The final temperature
reached by the gas after virialization depends on the mass of the object,
$M$, through the virial condition:
\begin{equation}
T_{\rm vir} \simeq  800 \, {\rm K} \left ({M \over 10^6 M_\odot} \right )^{2/3}
\left ({1+z \over 20}\right) \left ( { \Omega_m \over 0.3} \right )^{1/3}\left ( {h \over 0.7} \right )^{2/3}  \left ( { \mu  \over 1.22} \right ).
\label{vircon}
\end{equation}
\section{Primordial Magnetic fields and energy dissipation in 
the post-recombination era}
Next, we discuss the importance of primordial tangled magnetic field for
the production of molecular hydrogen.
Primordial tangled magnetic fields can play an important role in the 
formation of 
structures in the universe. In particular, they can lead to early formation of 
structures and leave detectable traces in the CMBR temperature and polarization
anisotropies (e.g. Barrow, Ferreira \& Silk
1997; Subramanian \& Barrow 1998a, 1998b; Durrer, Ferreira, 
Kahniashvili, 2000; Seshadri \& Subramanian 2001; Caprini \& Durrer 2002;  Subramanian, Seshadri \& Barrow 2003; Lewis 2004, Gopal \& Sethi 2003, 2005;
Yamazaki, Ichiki, \&  Kajino 2005,  Yamazaki et~al. 2006b,  Wasserman 1978, Giovannini 2006). 

Sethi and Subramanian (2005) 
studied the effects of the  dissipation of magnetic field energy owing to ambipolar diffusion and 
decaying turbulence in the post-recombination era. In particular they showed that
for magnetic field strengths $\simeq 3 \times 10^{-9} \, \rm G$
this dissipation can result in thermal and ionization history substantially different
from the usual case. 

Here, we also generalize the analysis of Sethi and Subramanian (2005) to take into
account the effects of this dissipation in a spherically collapsing halo. Our main
aim here is to study the effects of this dissipation on the formation of 
molecular hydrogen in the background IGM and also in a collapsing halo.
The magnetic field energy dissipation owing to ambipolar diffusion can be expressed as (see .e.g 
Cowling (1956), Shu (1991)):
 \begin{equation}
\left({dE_B \over dt}\right)_{\rm ambi} = {\rho_n \over 16 \pi^2 \gamma \rho_b^2 \rho_i} 
|({\bf \nabla x B) x B}|^2
\label{eq:amdif}
\end{equation} 
Here $\rho_n$, $\rho_b$, and $\rho_i$ correspond respectively to the 
neutral, total, and ionized mass density;  $\gamma = \langle w \sigma_{\rm in} \rangle /(m_n + m_i)$ (Shu 1992), 
where $w$ is the ion-neutral relative velocity and $\sigma_{\rm in}$ is the 
cross section for the collision between ions and neutrals. 
For $w \la 10 \, \rm km \, sec^{-1}$, 
$\langle w \sigma_{\rm in} \rangle \simeq 3 \times 10^{-9} \, \rm cm^3 \, sec^{-1}$ independent 
of the relative velocity of ions and neutrals. We use this value throughout
as it is a valid approximation in our case.
We assume the tangled magnetic field to be statistically homogeneous and isotropic and 
Gaussian with a power spectrum: $M(k) = A k^n$ with a large-k cut off
at $k = k_{\rm max}$; $k_{\rm max}$ is determined by the effects of damping 
by radiative  viscosity during the pre-recombination era (Jedamzik, Katalini{\' c}, \& Olinto 1998, Subramanian \& Barrow 1998b; for more details 
see  also Sethi and Subramanian 2005). The normalization $A$ can be determined by smoothing the 
magnetic field over a given scale $k_{\rm \scriptscriptstyle G}$. We continue 
to refer to the RMS of the tangled magnetic field smoothed over this 
scale (with a sharp k-space filter)  with $k_{\rm \scriptscriptstyle G} = 1 \, \rm Mpc^{-1}$ as the magnetic field
strength, $B_0$. 
We assume the magnetic field
power spectrum to be nearly scale invariant with, $n \simeq -3$,  for our study. Many theoretical analyzes  show that these are  the only power spectra compatible with
current observations (e.g. Caprini \& Durrer 2002, Sethi and Subramanian 2005 and references therein).
For obtaining numerical results, we use $n = -2.8$ throughout this paper.
Also, for these
spectra,  the RMS is not sensitive to the exact choice of the smoothing scale. 

In linear theory, it results in the flux freezing condition $Ba^2 = {\rm const}$, where 
$a$ is the scale factor of the expanding universe.  
Using this,  the magnetic field term in the 
above expression can be simplified to (for details see Sethi and Subramanian (2005)):
\begin{equation}
\left\langle \left[({\bf \nabla}_{\bf x} \times \tilde{\bf B}) \times  
\tilde{\bf B}\right]^2 \right \rangle = {7 \over 3} \int dk_1 
\int dk_2 M(k_1)  M(k_2) k_1^2 k_2^4
\label{eq:n6pp}
\end{equation}
Here $x$ refers to comoving coordinates and $\tilde{\bf B} = {\bf B} a^2$ is the value of the magnetic field at the present epoch. The expression given above 
is exact in linear perturbation theory. For taking into account the formation of 
non-linear structures, we continue to use the flux-freezing condition and generalize it 
to: $B n_b^{2/3} = {\rm const}$ and allow the comoving length (constant in 
linear theory)  to be $ \propto [n_b^{1/3}(1+z)]^{-1}$. 

Magnetic field energy  can also dissipate by generating decaying MHD turbulence. Sethi and Subramanian (2005) modelled this  dissipation owing to  MHD 
turbulence based on large scale simulation of this phenomenon in flat space. 
The simulations 
show that, the decay of magnetic field can be modelled as (see e.g. Banerjee \& Jedamzik 2003, 
Christensson, Hindmarsh \& Brandenburg 2001, 
Biskamp \& M{\"u}ller 2000): 
\begin{equation}
{\cal E}_{\rm B} = {{\cal E}_{\rm B0} \over (1 + \tilde t/\tilde t_{d})^m}.
\label{flatdec}
\end{equation}
Here $\tilde t$ is the time in 'flat-space',
$\tilde t_{d}$ is the relevant dynamical time for decay,
and ${\cal E}_{\rm B}$ is the magnetic energy in flat space,
with ${\cal E}_{\rm B0}$ its initial value. As discussed in Sethi and Subramanian (2005), 
the decaying turbulence sets in as  the radiative viscosity drops rapidly during 
the recombination era. For our computation,  we 
assume the decaying turbulence to set in  at $z = 1000$, and therefore the initial 
magnetic field energy ${\cal E}_{\rm B0}$ is computed at this redshif. 
A change in the choice of initial redshift  value doesn't make any essential difference to our results
as the initial phase of dissipation is always logarithmic (to be discussed below).  Simple
scaling arguments suggest $m = 2(n+3)/(n+5)$, for
an initial power spectrum with $n \ga  -3$ (cf. Olesen 1997, Shiromizu 1998,
Christensson, Hindmarsh \& Brandenburg 2001, 
Banerjee \& Jedamzik 2003). This flat space result can be mapped to the case of 
expanding universe by using the following transformation of MHD equations 
(Banerjee \& Jedamzik 2004):
\begin{equation}
\tilde{\bf B} = {\bf B} a^2, \quad 
d\tilde t = dt/a^{3/2}, \quad \tilde{v} = a^{1/2} v, \quad 
\tilde \rho = \rho a^3, \quad \tilde p = p a^4, \quad 
\tilde \epsilon = a^4 \epsilon
\label{trans}
\end{equation}
For our study, the most important transformation concerns the transformation of time. 
As was shown in Sethi and Subramanian (2005), this means that a power-law dissipation
in flat space translates to a logarithmic decay in the expanding universe.
 We generalize
this concept to a collapsing halo here. Following Sethi and Subramanian (2005),
the rate of dissipation of magnetic field energy owing to decaying turbulence 
can be expressed as:
\begin{equation}
\left({dE_B \over dt}\right)_{\rm turb} = m{B^2(t) \over 8 \pi} {d\tilde t \over dt} {1\over \tilde t_d} {1\over (1+\tilde t/\tilde t_d)}
\end{equation} 
For a collapsing halo, the transformation between $\tilde t$ and 
$t$ (Eq.~({\ref{trans})) can be computed by replacing the scale factor 
 by  the comoving length of the collapsing object. From Eq.~({\ref{trans}), 
 it can  readily be shown that the dissipation in a collapsing halo is generally a power law and  
the dissipation of the magnetic field energy in a collapsing
 halo is generally  more rapid.
The evolution of magnetic field energy density can be written as:
\begin{equation}
{dE_B \over dt} = {4 \over 3} {\dot \rho \over \rho}  - \left({dE_B \over dt}\right)_{\rm turb}  - \left({dE_B \over dt}\right)_{\rm ambi}
\end{equation}
The first term on the right hand side takes into account 
the change in the magnetic field energy owing to adiabatic expansion (in
the early stages of the evolution of a halo) and compression (during the 
halo collapse). The heating terms in Eq.~(\ref{temp}) correspond to:
\begin{equation}
L_{heat} = \left({dE_B \over dt}\right)_{\rm turb}  + \left({dE_B \over dt}\right)_{\rm ambi}
\end{equation}

\section{Results}
The ionization and thermal history for several values of magnetic  field
strengths are  shown in Figure~1~and~2 (the virial redshift $z_{\rm vir} = 10$ for 
all the figures in the paper unless specified otherwise).    The dissipation 
of magnetic field energy is
seen to have significant effect on the thermal and ionization history in the 
post-recombination era. 
The more interesting case is the 
case of a collapsing halo. In this case, the magnetic  field strength increases
owing to the flux freezing condition 
as the halo starts collapsing, which also leads to an
increase in the amount of 
magnetic field energy that is dissipated as more 
magnetic field energy becomes available. 
During the collapse phase, the heating mechanism 
of the halo is provided by both the adiabatic compression as well as 
the magnetic field energy dissipation. The main cooling mechanism for halos
that exceed  temperatures $\ga 10^4 $ K is the HI line cooling 
 and this prevents the halo to reach temperatures 
much higher than this value, as is also seen in Figure~2.

We do not include here the temperature increase owing to the process of
the virialization of the halo. This final increase in a collapsing halo
depends on the mass of the halo as  obtained by  the virialization condition (Eq.~(\ref{vircon}). 
The temperature histories we present for collapsing halos, however,
do not depend on the mass of the halo but only on the strength of the magnetic 
field. 
This requires one to interpret the virialization condition 
differently. In the usual case without magnetic fields, the temperature
reached by a collapsing halos (of  mass scales of interest $\ga 10^5 \, \rm M_\odot$) owing to 
adiabatic compression is generally much lower than given  by the 
virialization condition; this is because the final density is, as discussed above, roughly 
$18\pi^2$ times the  IGM density, which gives the final temperature owing 
to adiabatic compression: $T_f \simeq (18\pi^2)^{2/3}T_{\rm IGM}$. 
  In our case, however,  the temperature reached owing
to magnetic field energy  dissipation coupled to adiabatic collapse is far 
higher than this, and  it is 
quite comparable to virial temperature of halos of mass scales 
of interest.  

We therefore interpret the final temperature ($T_f$) reached by 
a halo to correspond to the virial temperature ($T_v (M)$) 
of the appropriate mass scale.
This means that  halos of smaller masses (with $T_v (M) < T_f$) 
cannot collapse 
gravitationally (or will evaporate during the collapse process)  for the 
corresponding value of the magnetic  field.  For larger 
masses, the virialization process can raise the temperature of the halo
above the value we obtain for a given value of magnetic field strength
($T_f > T_v (M)$). The  
line cooling, however,  prevents the  the temperature of the halo from
rising above a value  $\simeq 10^4 \, \rm {K}$, as seen in Figure~2. 

The virial temperature defined in this way is  related to the usual 
Jeans' criterion. Given the density and temperature evolution of the gas,
the Jeans' criterion can be expressed as a lower limit on the mass
that is unstable to gravitational collapse, with this lower limit given by:
\begin{equation}
M \ge M_J = {4 \pi^4 c_s^3 \over 3(4\pi G)^{3/2} \rho_m^{1/2}}
\label{jeanmass}
\end{equation}
Here $c_s^2 = (5/3) KT/(\mu m_p)$ is the square of sound speed at
constant entropy.

\subsection{Molecular hydrogen in the IGM}
The rate of molecular hydrogen formation is a fairly sensitive function
of the ionization fraction in the IGM (or collapsing halo) (see e.g. Tegmark 
et~al. 1997). Owing to the dissipation of magnetic field energy, 
the IGM temperature is much higher than in the usual case, and consequently, the
ionization fraction in the IGM is much higher than the usual case, as 
seen in figure~1 and~2.  
The rate of molecular hydrogen formation increases as ionization fraction
increases. However, a rise in temperature also increases the rate of destruction
of $H^-$ which is the intermediate product for the formation of the molecular 
hydrogen. It is an interplay of these two opposing effects that determines
the production of molecular hydrogen in the presence of primordial 
magnetic field energy  dissipation. 

In Figure~3, we show the evolution of molecular hydrogen formation 
for several values  of the  primordial magnetic field. As seen in the Figure, 
the molecular hydrogen fraction is generally seen to increase with the 
strength of magnetic field. This shows that of the two opposing influences,
the rise in the ionization level owing to magnetic field energy  dissipation
 predominates in determining the production rate of molecular hydrogen. 
The dissipation of tangled magnetic field energy  in the post-recombination era
can, therefore,  play a very important role in determining the formation of molecular
hydrogen. As seen in Figure~3, the fraction of molecular hydrogen 
in the IGM can increase by more than two orders of magnitude owing to
this process, for magnetic field strength $\simeq 3 \times 10^{-9} \, \rm G$. 
This can have several important implications for the 
formation of structures in the early universe and also the radiation
transfer of UV light in the early universe. We discuss it in more detail
in a later section. 

\subsection{Molecular hydrogen in collapsing haloes}
In Figure~3 we show the molecular hydrogen fraction in collapsing haloes
for many different values  of the magnetic field. It is seen that, 
as in the usual
case without magnetic field, the fraction of molecular hydrogen
 increases rapidly during the final  collapse phase of the halo. 
However, as Figure~3 shows, the increase in 
molecular hydrogen fraction in the pre-collapse phase 
in the presence of magnetic fields is  also quite important and  
might  be primarily responsible for the increase in
 the fraction of molecular hydrogen
as the  magnetic field value is increased.  In addition, as seen in Figure~1, 
the final collapse phase could be accompanied by an increase in the 
ionized fraction in the presence of magnetic fields, unlike the usual case; 
this leads to a further increase in the molecular hydrogen fraction. 
This  suggests that  the 
increase in molecular hydrogen fraction is determined by 
the  entire   ionization
and thermal history in the presence of magnetic fields. 
  For magnetic fields strengths of 
$B_0\ga 2.5 \times 10^{-10} \, \rm G$, the final 
abundance of molecular hydrogen fraction 
could reach values $\simeq 10^{-3}\hbox{--}10^{-2}$. 
\section{Discussion}
The possibility of increased abundance of H$_2$ in the IGM and 
the collapsing halos has important
implications for  the evolution of luminous structures in the universe.

In the usual case without magnetic fields, the smallest mass haloes (above
the thermal Jeans mass $\simeq 10^5 \, \rm M_\odot$) that satisfy 
the criterion for star formation ($t_{\rm cool} \le t_{\rm dyn}$) 
can only cool by molecular hydrogen (e.g. Tegmark et al. 1997, for 
a review see Barkana \& Loeb 2001). These first haloes are generally
believed to have masses $\ga 10^7 \, \rm M_\odot$. However, there 
 are two difficulties in forming and sustaining star-formation using 
$\rm H_2$-cooled haloes: (a) the cooling time $t_{\rm cool}$
depends directly on the abundance of molecular hydrogen fraction. 
In the usual case the fraction of molecular hydrogen is small $\simeq 2 \times 10^{-4}$, which barely meets the condition for runaway collapse of 
haloes owing to $\rm H_2$ cooling. (b) Even if the first stars  could form,
the UV light from these first object can easily penetrate the surrounding
IGM and destroy the $\rm H_2$ in the neighboring haloes (Haiman, Rees, \& Loeb  1997) 
thereby inhibiting further star-formation. The scenario in the 
presence of magnetic fields can obviate both these difficulties but 
it also introduces additional complications. 

The introduction of magnetic field energy dissipation is seen to increase the 
molecular hydrogen fraction by up to two orders of magnitude (Figure~3). 
This decreases  the cooling time of the collapsing halo and would seem
to suggest that haloes would not 
only collapse more readily but haloes of  a wider 
set of masses would now become unstable to collapse. However, this is 
not the case owing to two reasons: (a) as discussed above, the temperature
reached in the collapsing haloes are higher owing to the magnetic field energy
dissipation. If this temperature exceeds the thermal Jeans' 
mass (Eq.~(\ref{jeanmass}) then the halo cannot collapse, (b) haloes of 
masses smaller than magnetic Jean's mass cannot collapse (for details
of structure formation in the presence of magnetic fields see e.g. Sethi
and Subramanian 2005 and discussion below). 

In Figure~4, we show the minimum thermal Jeans' mass  (obtained from
the condition given by  Eq.~(\ref{jeanmass})) and the magnetic Jeans' mass
as a function of the magnetic field strength. 
The magnetic Jeans' mass is given by  (see .e.g. Sethi \& Subramanian 2005):
\begin{equation}
M_J^B \simeq  10^{10} \, {\rm M_\odot} \left({B_0 \over 3 \times 10^{-9} \, {\rm G}} \right )^3. 
\end{equation} 
The  haloes of masses 
that lie in the range  above both
the curves shown in the figure can  gravitationally collapse. 

Some notable features of Figure~4 and their 
implications are: (a) magnetic Jeans's mass
exceeds the thermal Jeans' mass for magnetic field strengths 
 $\la 10^{-10} \, \rm G$ and $\ga 10^{-9} \, \rm G$. (b) 
there exists a mass range $10^6 \la M \la 10^8 \, \rm M_\odot$, which
is Jeans' unstable according to both thermal and magnetic criteria. 
 For a part of this mass range 
the magnetic field strength is between roughly $2\hbox{--}4 \times 10^{-10} \, \rm G$. For these magnetic field strengths,  as follows from Figure~3, 
the molecular hydrogen fraction  in the collapsing halos 
 could be in the range $1\hbox{--}5 \times 10^{-3}$, which is more than
ten times higher than the usual case (see e.g. Tegmark et~al. 1996).
Therefore, the net effect of primordial magnetic field is to increase
the molecular hydrogen fraction in an interesting range of masses.
And the main implication of this increase is  to  increase the 
cooling efficiency and therefore the rate of collapse of these  haloes.
It also follows from Figure~3 and~4, that haloes with higher molecular
hydrogen ($\ga 0.01$) have masses $> 10^8 \, \rm M_\odot$; these 
haloes, inspite of having higher molecular fraction, cool faster
by atomic cooling (e.g. Barkana \& Loeb 2001, Sethi 2005). 

  Sethi
and Subramanian (2005) showed that primordial magnetic fields can induce early 
formation of structures. In particular they showed that for 
magnetic-field induced structure formation: 
(i) the mass dispersion $\sigma_B(R)$ could reach values 
large enough to cause a 1-$\sigma$ collapse at high redshifts;  the
value of collapse redshift 
is only sensitive to the magnetic field spectral index and 
not its strength
(provided $M > M_J$)
(ii) $\sigma_B(R) \propto 1/R^2$, or the mass dispersion
 decreases quite rapidly with the 
mass scale and therefore most of collapsed structures would have 
masses close to the magnetic Jeans' mass. The magnetically-induced 
structure formation is in addition to the 
structure formation  in the  usual $\Lambda$CDM model.  
The mass dispersion at any scale in the presence of 
primordial magnetic field is: $\sigma(R) = \sqrt{\sigma_{\Lambda CDM}^2 + \sigma_B^2}$.   These conclusions can be 
suitably revised in the light of results shown in Figure~4. 

 Sethi
and Subramanian (2005) (Figure~7) showed that for magnetic field spectral
index $n = -2.8$, $\sigma_B \simeq 1.5$ at the scale of magnetic Jeans' length,  at $z \simeq 10$. This means that at this redshift a 1-$\sigma$ 
density perturbation
can collapse. It follows
from Figure~4, that for magnetic field strength $B_0\la 10^{-9}$, 
it is usually the thermal Jeans' mass that is important for the collapse 
criterion. For example, for $B_0 \simeq  4 \times 10^{-10} \, \rm G$, the 
thermal Jeans' mass is roughly a factor of a factor of 5 greater than 
the magnetic Jeans' mass; the mass dispersion at this scale is roughly
$0.8$, which means only a 2-$\sigma$ perturbation can collapse. This 
is in addition to $\Lambda CDM$ model-induced perturbations. 
The collapse of these perturbations in not changed
 much by the increase in Jeans' length shown in Figure~4 
because  $\sigma_{\Lambda CDM}$ is roughly independent of scale at
small scales (scales corresponding to $M \la 10^8 \, \rm M_\odot$). 
From WMAP-normalized $\sigma_{\Lambda CDM}$ power spectrum, $\sigma_{\Lambda CDM}$ at scales under discussion is $\simeq 0.7$, which means a 2.5-$\sigma$
perturbation can result in collapsed structures. 

For magnetic field spectral
index $n = -2.9$, $\sigma_B \simeq 0.5$ at the scale of magnetic Jeans' length,  at $z \simeq 10$ (Figure~7, Sethi
and Subramanian 2005).  This means magnetic field-induced 
structure formation is not as dominant as  the  $\Lambda CDM$ model.
 The thermal history of the universe   
owing to magnetic field energy dissipation and its implications, as presented here,  
 are   not sensitive to the 
value of $n$. Therefore,  the main effect of the presence of 
primordial magnetic fields for  $n \la -2.9$ is to alter the thermal
history and the $H_2$ formation in the universe. However,
magnetically-induced structure formation would also be important
for larger values of $n$.

As mentioned above, the formation of structures with $\rm H_2$-cooled 
haloes is self-inhibiting owing to the destruction of $\rm H_2$ by the  
UV photons emitted  by first objects. This is because IGM is optically 
thin to 
the UV photons in the Larmer-Werner (LW) band ($11.8\hbox{--}13.6$ eV). 
 In the
usual case,  the abundance of $\rm H_2$ in the IGM is   
$\simeq  10^{-6}$ and the  photons emitted by
first generation (Population III) stars in the 
Lyman-Werner  band 
 are absorbed to some extent by the $\rm H_2$ molecule in the IGM.
These photons, if unabsorbed, can photodissociate $H_2$ molecule in 
neighbouring objects. This phenomenon may  cause a decrease in the
star formation rate in the universe, until more massive objects form through 
atomic cooling begin to appear in abundance, or enough $\rm H_2$ molecule
has form by the influence of ionization from X-ray photons which may
be  present (Haiman, Abel, Rees 2000; Glover, Brand 2003). 

Haiman, Abel and Rees (2000) estimated that for an
abundance of $\simeq 2 \times 10^{-6}$ in the IGM, the opacity of LW
band photons can be as large as $\simeq 0.2$  at $z \simeq
15$. As discussed by Cen (2003), the efficacy of LW photons in dissociating
H$_2$ molecule--- which competes with production of H$_2$ molecules with the
help of free electrons generated by higher energy photons--- can be
described by a ratio $\psi \sim E_{LW}/E_{>1 \, {\rm keV}}$. Consider 
a background radiation with spectrum $J_\nu \propto \nu^{-1}$, as considered
by Haiman et al (2000). Cen (2003)
summarized the results of Haiman et al (2000) (which had likely underestimated
the opacity of LW photons by a factor $\simeq 6$ (Cen 2003)) to state  that the destruction
of $\rm H_2$ molecules by LW bands is compensated by production of $\rm H_2$ molecules
owing to ionization by higher energy photons, if $\psi \le 5$. Typically,
for Population III stars
Cen (2003) estimated the ratio to be $\psi \sim 2.5$. 

  The opacity of LW photons scales linearly with the H$_2$ molecule abundance.
 Therefore, if the IGM abundance of H$_2$ molecule increases even by a factor
$\simeq 5\hbox{--} 30$ (which corresponds to a magnetic field of order $\simeq 4 \times 10^{-10} \, \rm G$, from
Figure 3), the opacity of LW band photons can be large enough to block
them from destroying H$_2$ molecule in nearby objects. In this case, the
star formation history is likely to be less affected by the UV emission of 
Population III stars than in the usual case. Therefore, the presence
of tangled primordial magnetic field in the IGM  can have substantial
effect in the history of star formation in the universe. 

\section{Summary and Conclusions}
In this paper, we explored the implications of the presence of primordial
magnetic fields on the formation of molecular hydrogen in the first
collapsed objects in the universe. 

We showed that the dissipation of the primordial magnetic field energy 
 in the post-recombination era can alter the ionization and thermal evolution substantially.
The net effect of this change is  
to increase the molecular hydrogen abundance by many orders of magnitude,
depending on the value of the magnetic field. 

We discuss how the 
increase in the molecular hydrogen abundance can have a direct bearing 
on the formation and survival of the first objects. The net impact 
of the primordial magnetic fields depends also on the role played 
by the magnetic fields in inducing the formation of first structures. 

In future, we hope to explore the observational consequences of the 
effects of primordial magnetic fields in the formation of the first structures
in the universe.

\section*{Acknowledgment}
We  would like to thank Zoltan Haiman for 
providing data on the absorption of UV photons in the Larmer-Werner band. 
We also thank S. Sridhar for  fruitful
discussion.

\begin{figure}
\epsfig{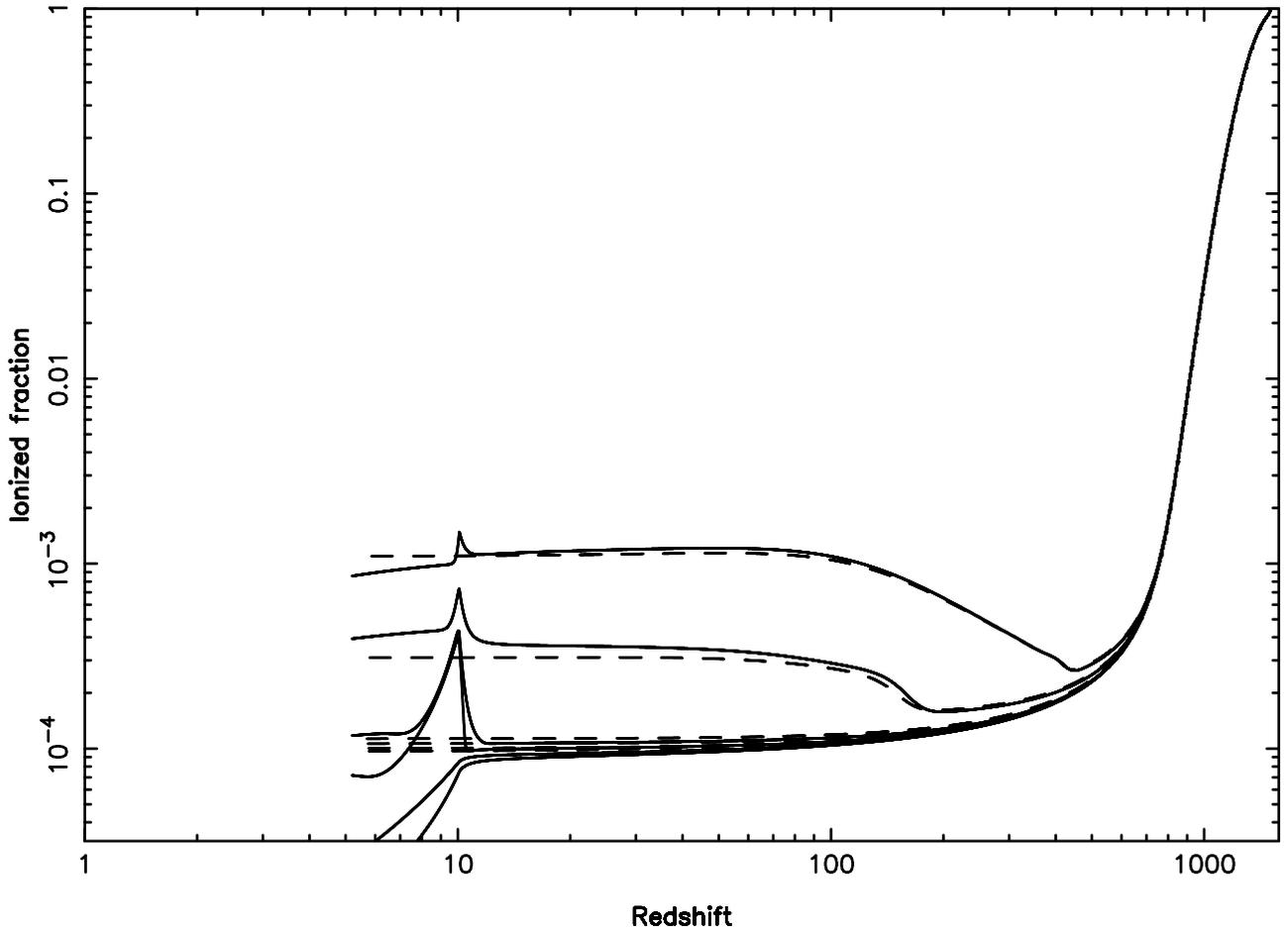}
\caption{The ionized history is shown for the IGM (dashed lines) and a collapsing 
haloes (solid curves) for several different choices of the magnetic field strength. The 
different curves (from bottom up) correspond to magnetic field strengths: $\{10^{-10}, 2.5 \times 10^{-10}, 3.5 \times 10^{-10}, 6 \times 10^{-10}, 10^{-9}, 3 \times 10^{-9} \} \, \rm G$.  }
\label{fig:f1}
\end{figure}

\begin{figure}
\epsfig{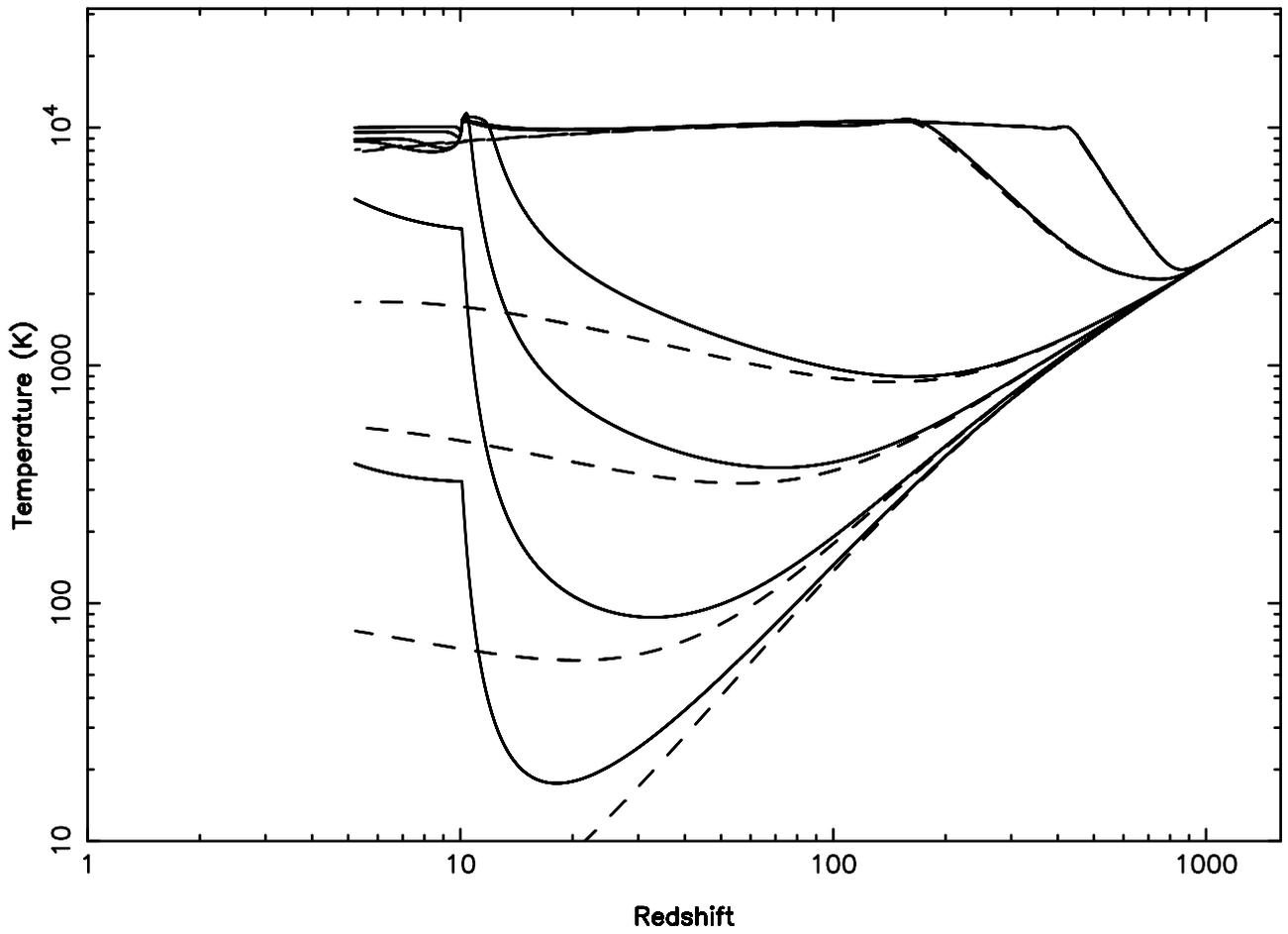}
\caption{The thermal history of IGM and collapsing halo is shown. The parameters
and the plotting style is the same as Figure~1 }
\label{fig:f2}
\end{figure}

\begin{figure}
\epsfig{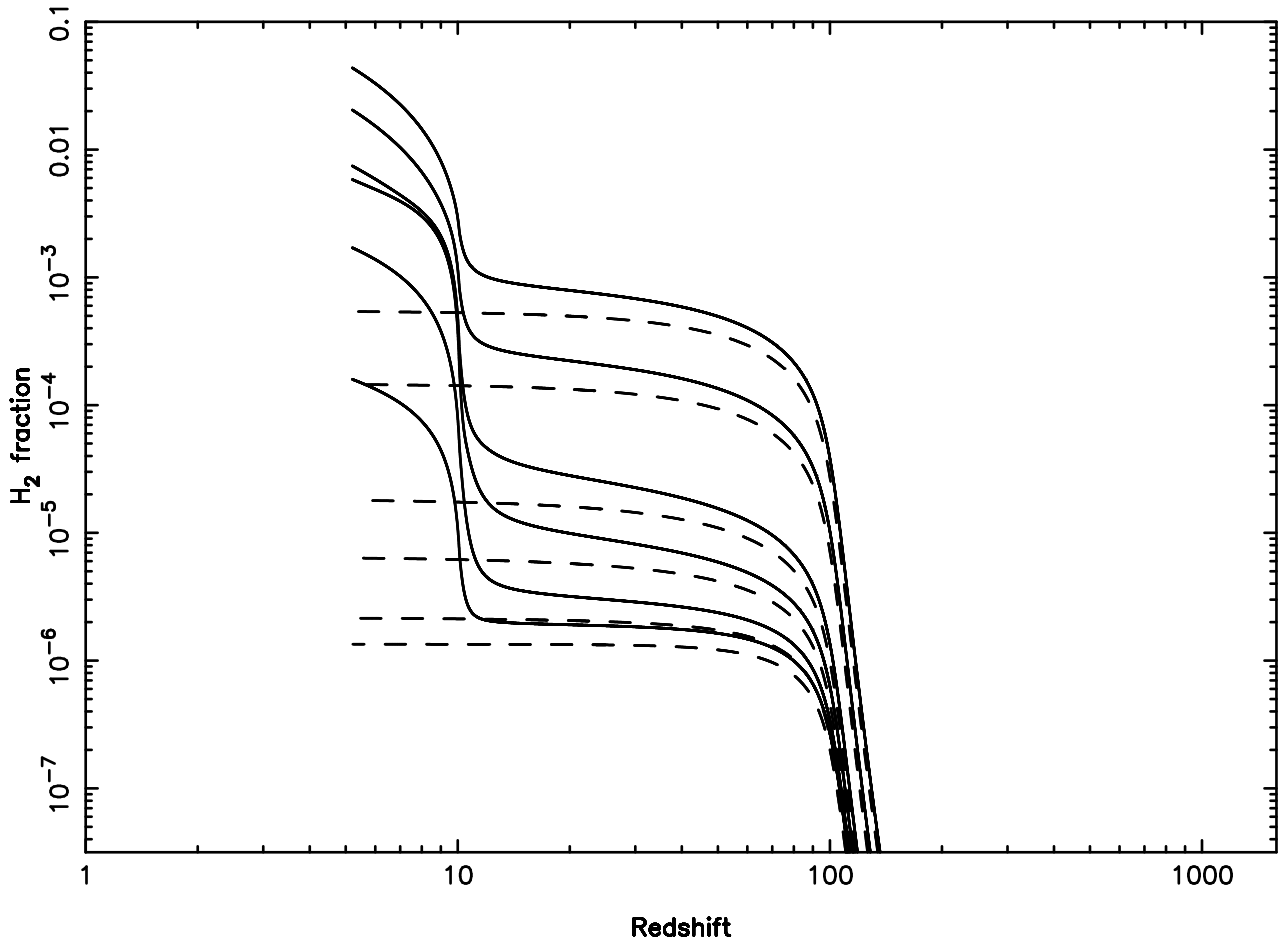}
\caption{The molecular hydrogen fraction in  IGM and collapsing halos is shown. The parameters
and the plotting style is the same as Figure~1 }
\label{fig:f3}
\end{figure}

\begin{figure}
\epsfig{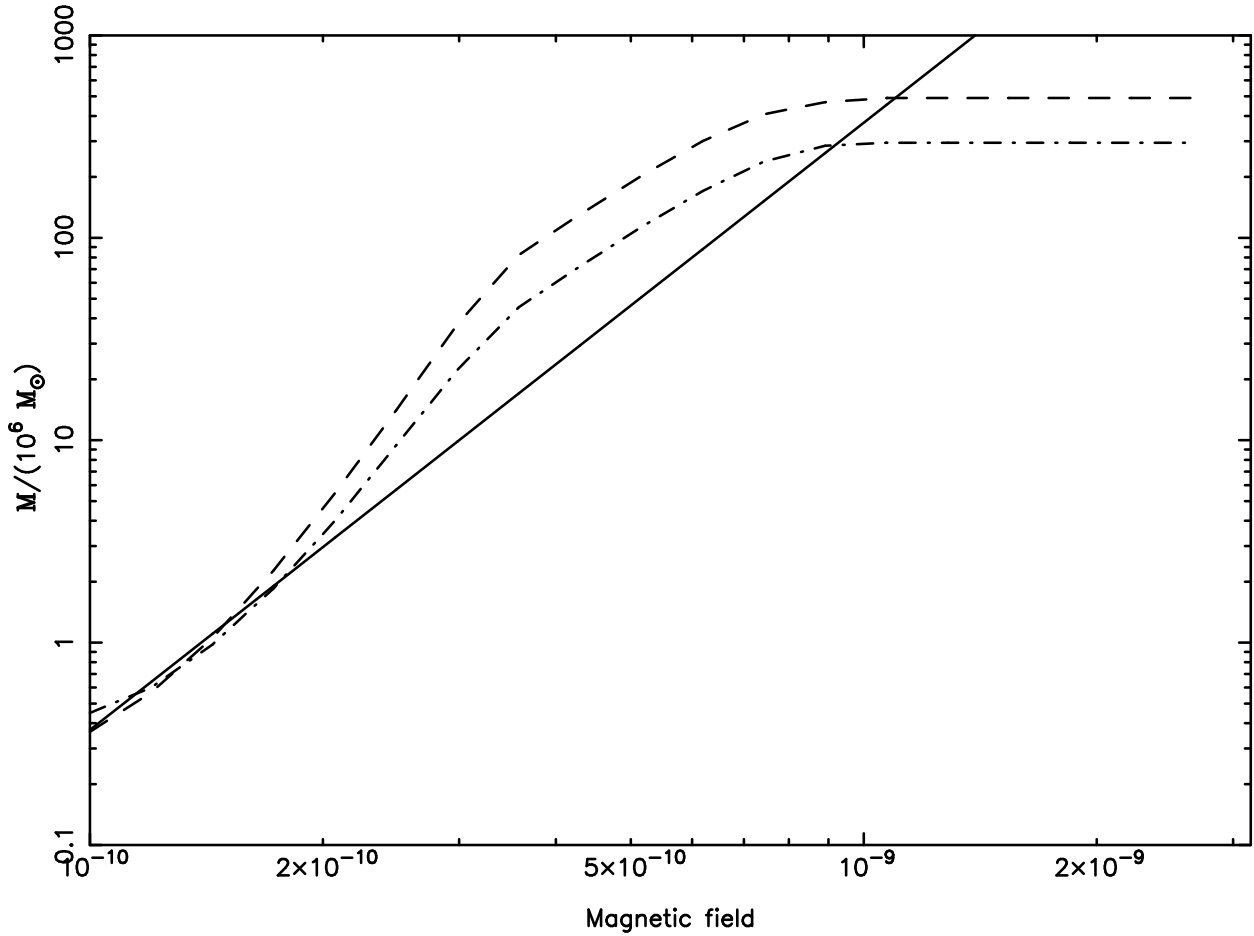}
\caption{The solid curve shows the magnetic Jeans mass as a function of 
Magnetic field strength. The dashed   curves, top to bottom, are obtained by the 
condition given by Eq.~(\ref{jeanmass}) for $z_{\rm vir} = \{10,15\}$, 
respectively.}
\label{fig:f4}
\end{figure}

\end{document}